\documentclass[12pt]{article}
\usepackage{graphicx}

\usepackage{scicite}

\usepackage{times}

\newenvironment{sciabstract}{%
\begin{quote} \bf}
{\end{quote}}

\newcounter{lastnote}
\newenvironment{scilastnote}{%
\setcounter{lastnote}{\value{enumiv}}%
\addtocounter{lastnote}{+1}%
\begin{list}%
{\arabic{lastnote}.}
{\setlength{\leftmargin}{.22in}}
{\setlength{\labelsep}{.5em}}}
{\end{list}}

\title{A gaseous metal disk around a white dwarf}

\author{B.T. G\"ansicke$^1$, T.R. Marsh$^1$, J. Southworth$^1$,
A. Rebassa-Mansergas$^1$\\
\normalsize{$^{1}$ Department of Physics, University of Warwick, Coventry CV4
   7AL, UK}\\
}

\date{}

\begin{document} 


\baselineskip24pt


\maketitle



\begin{sciabstract}
The destiny of planetary systems through the late evolution of their
host stars is very uncertain. We report a metal-rich gas disk around a
moderately hot and young white dwarf.  A dynamical model of the
double-peaked emission lines constrains the outer disk radius to just
1.2 solar radii. The likely origin of the disk is a tidally disrupted
asteroid, which has been destabilised from its initial orbit at a
distance of more than 1000 solar radii by the interaction with a
relatively massive planetesimal object or a planet. The white dwarf
mass of 0.77 solar masses implies that planetary systems may form around
high-mass stars.
\end{sciabstract}


White dwarfs are the compact end products of stars with masses up to
$\sim8$ solar masses \cite{dobbieetal06-1}.  Because of the low
luminosity of white dwarfs the detection of low mass stellar
companions \cite{farihietal05-1} or planets \cite{burleighetal02-1} is
much easier around white dwarfs than around main sequence stars.
During a search for cool companions to white dwarfs, an infrared
excess was discovered around the white dwarf
G29--38\cite{zuckerman+becklin87-1}. The atmosphere of G29--38 has
been found to be enriched with metals, i.e. elements heavier than
helium. The sedimentation time scales of heavy elements in the
high-gravity atmospheres of white dwarfs are short compared to the
evolutionary time scale of these stars\cite{paquetteetal86-1} and,
hence, the high metal abundances in G29--38 imply that this star is
accreting at a relatively high rate\cite{koesteretal97-1}.  Deep
imaging and asteroseismological studies of G29--38 ruled out a brown
dwarf companion\cite{grahametal90-1, kuchneretal98-1} and led to the
hypothesis of a cool dust cloud around the white dwarf. The presence
of dust near G29--38 has been verified by infrared observations with
the Spitzer Space Telescope\cite{reachetal05-1}. Infrared surveys of
white dwarfs exhibiting metal enriched atmospheres recently led to the
discovery of three other potential dust disks\cite{becklinetal05-1,
kilicetal05-1, kilicetal06-1}.  A possible origin of such dust disks
is the tidal disruption of either comets\cite{debesetal02-1} or
asteroids\cite{jura03-1}. Asteroids appear to be more likely
candidates as they can explain the large amount of metals accreted by
the white dwarfs from the dusty environment as well as the absence of
hydrogen or helium.  While the detection of asteroid debris around
G29--38 and the other white dwarfs represents a possible link to the
existence of planetary systems around their main-sequence progenitors
stars, modelling the excess infrared luminosity provides no direct
information on the geometric location and extension of the dust,
impeding a more detailed understanding of the nature and origin of the
circumstellar material \cite{reachetal05-1}. A concentration of dust
in the equatorial plane around G29--38 has been suggested on the basis
of the relative amplitudes of non-radial white dwarf pulsations
observed in the optical and infrared wavebands \cite{grahametal90-1}.

We identified SDSS\,J122859.93+104032.9 as a moderately hot white
dwarf in the Data Release 4 of the Sloan Digital Sky Survey
(SDSS)\cite{adelman-mccarthyetal06-1}, but noted very unusual emission
lines of the Ca\,II 850-866\,nm triplet, as well as weaker emission
lines of Fe\,II at 502\,nm and 517\,nm. The line profiles of the
Ca\,II triplet display a distinct double-peaked morphology, which is
the common hallmark of a gaseous, rotating disk\cite{youngetal81-3,
horne+marsh86-1}.  Time-resolved spectroscopy (Fig.\,1) and photometry
do not reveal any radial velocity or brightness variations. These
data exclude the possibility that SDSS\,1228+1040 is an interacting
white dwarf binary, in which an accretion disk around the white dwarf
forms from material supplied by a nearby companion star. Furthermore,
the absence of Balmer and helium emission lines implies that the
gaseous disk around SDSS\,1228+1040 must be extremely deficient in
volatile elements, which independently rules out an interacting binary
nature for this object.

Our detection of double-peaked metal emission lines from a
circumstellar disk in SDSS\,1228+1040 provides direct evidence for
hydrogen and helium depleted material rotating around a white dwarf at
a very short distance in a flat disk-like structure.  Assuming that
the Ca\,II line profiles (Fig.\,1) originate indeed in a circumstellar
disk, it must have an azimuthal asymmetry to match the asymmetry in
the profiles.  Such asymmetries are known in the disks around B-type
emission line stars (Be stars)\cite{meillandetal06-1}.  Drawing upon
this analogy, we developed a dynamical model of the Ca\,II line
profiles (Fig.\,3, SOM Sect.\,1), which provides a robust constraint
on the outer radius of the disk of $\simeq1.2$ solar radii.

As the main-sequence stars hosting planetary systems evolve through
the red giant stage, they swell up in radius and destroy planets and
asteroids out to many hundred solar radii\cite{sackmannetal93-1}. The
white dwarf mass of SDSS\,1228+1040 implies a relatively massive main
sequence progenitor of $\sim4-5$ solar masses \cite{bloecker95-1},
which will have expanded to a radius of $\sim1000$ solar
radii\cite{hurleyetal00-1}.  It is therefore impossible that the
material making up the present-day disk has survived the giant phase
at its current location, and must instead have been brought inwards
from outside a distance of 1000 solar radii. Planetary debris which
migrated outwards to large radii during the giant phase is expected to
have relatively stable orbits unless perturbed by larger-mass
objects\cite{debesetal02-1}. A likely scenario is therefore that one
or more planets which survived the evolution of the progenitor of
SDSS\,1228+1040 destabilised the orbit of an asteroid at some point
after the end of the planetary nebula phase. Getting close enough to
the compact star, the asteroid is tidally disrupted, forming a disk of
metal-rich debris which subsequently sublimates in the radiation field
of the white dwarf. The radius derived from our dynamical model is in
fact compatible with the tidal disruption radius for a rocky
asteroid\cite{davidsson99-1}.

A strong Mg\,II 448\,nm absorption line is detected in the spectrum of
the white dwarf and implies a magnesium abundance in its atmosphere
comparable to that of the Sun. This is extremely unusual for white
dwarfs, which typically have pristine hydrogen atmospheres. The large
abundance of magnesium can only be explained by sustained accretion,
as the diffusion time scale for magnesium in a white dwarf atmosphere
of 22\,000\,K (Fig.\,2, Table\,1) is only
$\sim5$\,days\cite{koester+wilken06-1}.  Moreover, the accreted
material must be of very low helium abundance, as already small traces
of helium mixed into the radiative atmosphere of the white dwarf would
cause noticeable He\,I 447\,nm absorption, which is not observed.  The
absence of helium lines in the white dwarf spectrum places
an upper limit on the helium abundance of the material accreted from
the circumstellar disc of 0.1 times the solar value, which is an
independent evidence for a metal-rich composition of the disk around
SDSS\,1228+1040. 

No infrared excess has been found around metal-polluted white dwarfs
hotter than $15\,000$\,K\cite{kilicetal06-1}, which suggests that the
radiation field of these hot white dwarfs causes sublimation of a dust
disk. The case of SDSS\,1228+1040 demonstrates that planetary debris
material can be detected around younger and hotter white dwarfs in the
form of gaseous disks.  Prompted by the discovery of SDSS\,1228+1040,
we have inspected 406 SDSS spectra of white dwarfs with hydrogen
dominated atmospheres brighter than $g=17.5$ that are contained in the
SDSS Data Release\,4 \cite{eisensteinetal06-1}, and find just one
additional object which potentially exhibits flux excess in the region
of the Ca\,II triplet (SDSSJ104341.53+085558.2, Fig.\,S1), so
SDSS\,1228+1040 is clearly a rare object. The detection of a
metal-rich debris disk around this relatively massive white dwarf
indicates that the formation of planetary systems can take place also
around short-lived massive stars.

\bibliographystyle{Science}


\begin{scilastnote}
\item Acknowledgements BTG, TRM, JS, and ARM have been supported by 
PPARC in form of an Advanced Fellowship, a Senior Research
Fellowship, a postdoctoral grant, and a joint PPARC-IAC studentship. The
observations were obtained at the Spanish Observatorio del Roque de
los Muchachos of the Instituto de Astrof{\'\i}sica de Canarias using
the William Herschel Telescope and Isaac Newton Telescope. We thank
Peter Wheatley for constructive criticism on the manuscript, Detlev
Koester for discussions on diffusion time scales, and Ivan Hubeny for
his ongoing support of the TLUSTY/SYNSPEC codes.

Funding for the SDSS and SDSS-II has been provided by the Alfred
P. Sloan Foundation, the Participating Institutions, the National
Science Foundation, the U.S. Department of Energy, the National
Aeronautics and Space Administration, the Japanese Monbukagakusho, the
Max Planck Society, and the Higher Education Funding Council for
England. The SDSS Web Site is http://www.sdss.org/.

\end{scilastnote}


\clearpage

\centerline{\includegraphics[width=14cm]{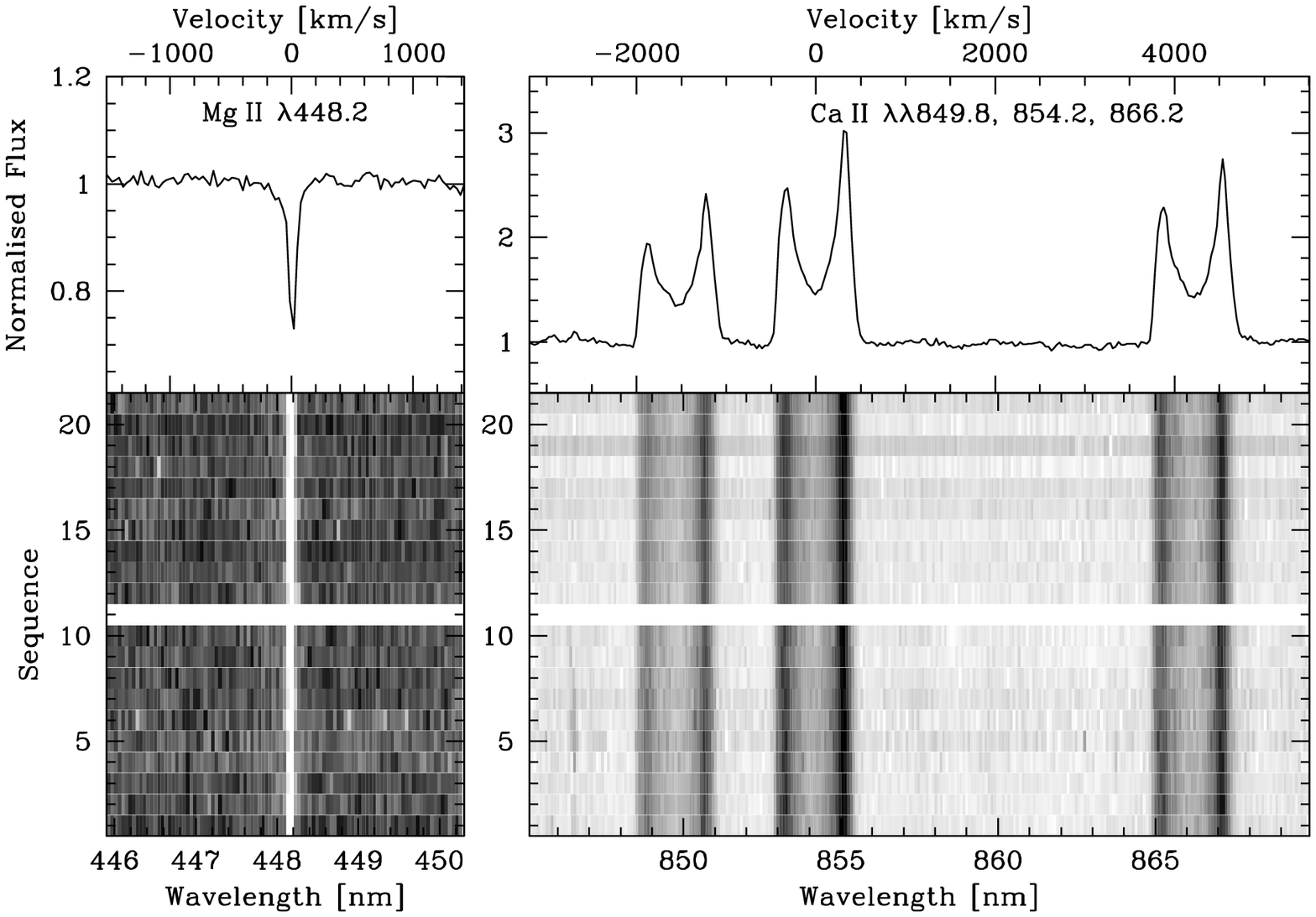}}

\medskip
\noindent {\bf Fig. 1. Time-resolved spectroscopy of SDSS\,1228+1040.}
Medium resolution spectroscopy of SDSS\,1228+1040 was obtained with
the double-arm spectrograph ISIS on the 4.2m William Herschel
Telescope on La Palma. Two sets of ten consecutive spectra each were
obtained on June 30 2006 and July 1 2006. The exposure times of the
individual spectra were 600\,s.  The bottom panel shows the two
time-series of spectra, each extending over 1.7\,h, centered on the
calcium emission triplet (right) and the magnesium absorption line
(left).  The normalised average spectra are shown in the top
panels. Radial velocities are given in the upper axes with respect to
Ca\,II 854\,nm (left) and Mg\,II 448\,nm (right). The radial velocity
of the Mg\,II line is stable to within $\pm4$\,km/s over time scales
of 20\,min to one day. Additional time-series photometry obtained at
the Isaac Newton Telescope shows the star at constant brightness
within $\pm0.01$\,mag.

\clearpage

\centerline{\includegraphics[angle=-90,width=11.1cm]{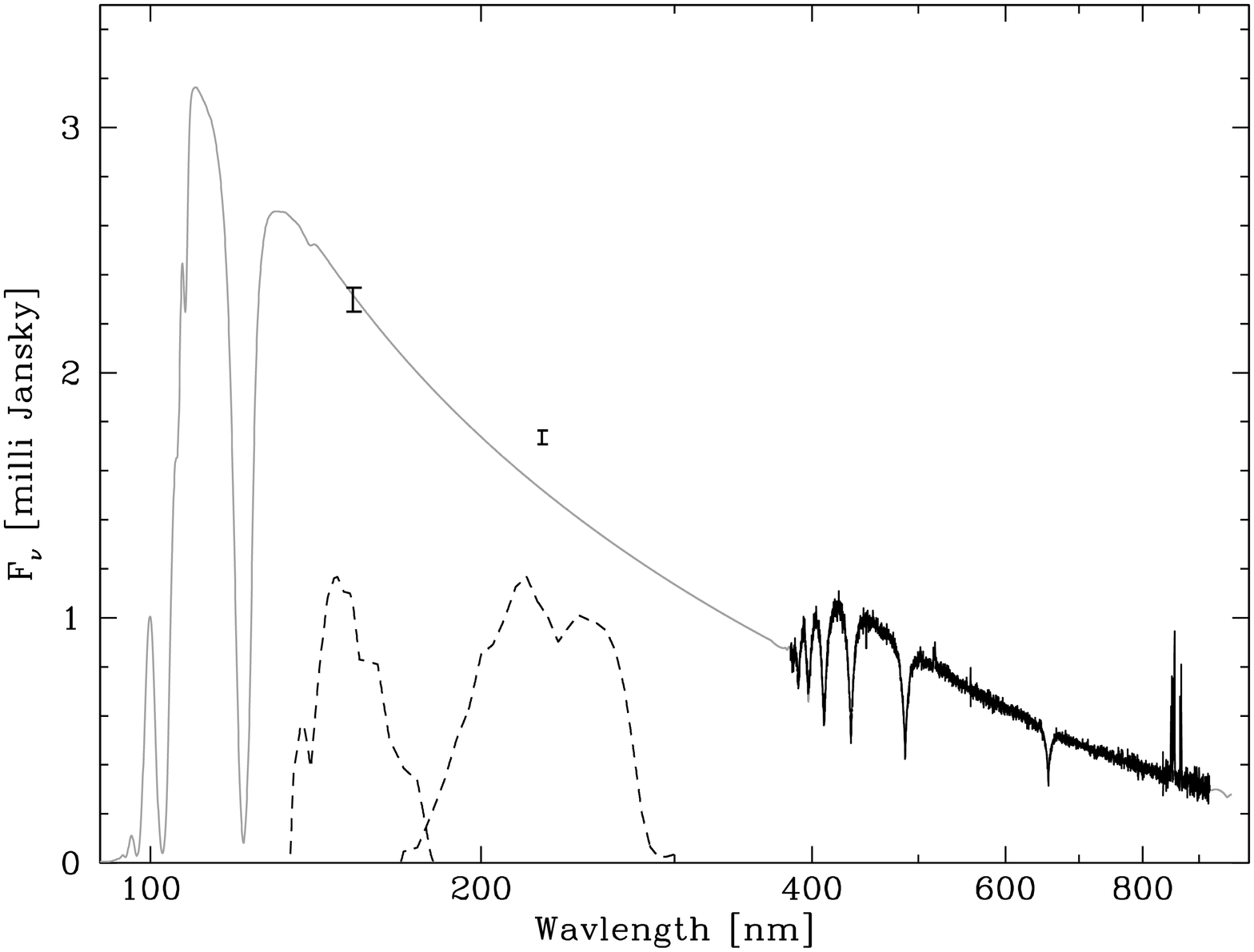}}

\medskip
\noindent {\bf Fig. 2. A model of the white dwarf in SDSS\,1228+1040.}
The optical SDSS spectrum of SDSS\,1228+1040 is plotted in black.  In
addition to the Ca\,II emission lines seen in Fig.\,1, also Fe\,II
517\,nm (and 502\,nm, not visible at the plotted scale) is
detected. The signal-to-noise ratio of the iron lines is too low to
assess the shape of their profiles.  The insets display blow-ups of the
H$\beta$ 486\,nm, H$\gamma$ 434\,nm, and H$\delta$ 410\,nm Balmer
lines and the Mg\,II 448\,nm absorption line.  The observed spectrum
has been modelled with synthetic white dwarf
spectra\cite{hubeny+lanz95-1}, resulting in an effective temperature
of $T_\mathrm{WD}=22\,020\pm120$\,K, a surface gravity of $\log
g=8.24\pm0.04$, an abundance of Mg\,II of $0.8\pm0.15$ with respect to
solar abundances, and a limit of $v \sin i\le20$\,km/s on the rotation
rate of the white dwarf. The best-fit white dwarf model is plotted in
gray. The ultraviolet fluxes detected by GALEX
(http://galex.stsci.edu/GR2/) are indicated as black error bars, the
filter transmission curves of GALEX are shown by dashed lines. The
best-fit white dwarf model is consistent with the far-ultraviolet
GALEX flux, but a flux excess is observed in the near-ultraviolet. A
host of Fe\,II transitions lies within the near-ultraviolet
bandpass\cite{koesteretal97-1}, and could be in emission in
SDSS\,1228+1040, explaining the observed flux excess.

\centerline{\includegraphics[angle=-90,width=14cm]{1135033Fig3.ps}}

\medskip
\noindent {\bf Fig. 3. A dynamical debris disk model.}  We modelled
the observed Ca\,II emission line profiles (points) by assuming that
the orbits in the disk took the form of a series of co-aligned
elliptical orbits of constant eccentricity (see SOM Sect.\,1). We
find an outer radius of 1.2 solar radii for an edge-on disk and an
ellipticity of 0.021.  The measured outer radius of the disk is
relatively invulnerable to the detailed assumptions of the model as it
is primarily fixed by the velocities of the emission line peaks. The
value given has to be interpreted as an upper limit since the radius
scales as $\sin^2 i$ for disks inclined by angle $i$ to the line of
sight. The deep central dips of the profiles imply optically thick
emission\cite{horne+marsh86-1} and suggest that the orbital
inclination must be quite high, $>70^\circ$, and therefore the $1.2$
solar radii upper limit should be close to the true value.

\clearpage

\medskip
\noindent {\bf Table 1. System parameter of SDSS\,1228+1040.}\\

\begin{tabular}{lr}
\hline
Distance                               &   142$\pm$15\,pc         \\
Effective temperature                  &  22020$\pm$200\,K        \\
White dwarf mass                       &   $0.77\pm0.02\,\mathrm{M}_\odot$ \\
White dwarf radius                     & $0.0111\pm0.003\,\mathrm{R}_\odot$ \\
White dwarf cooling age                &    100$\pm5$\,Myr        \\
Helium abundance                       & $\le0.1\times$solar      \\
Calcium abundance                      & $0.8\pm0.15\times$solar  \\
Progenitor mass                        &   $4-5\,\mathrm{M}_\odot$         \\
Progenitor main sequence life time     &   $\sim70$\,Myr          \\
Ca\,II full-width at zero intensity    &   $1270\pm35$\,km/s      \\
Ca\,II peak separation                 &   $630\pm5$\,km/s        \\
Outer radius of the circumstellar disk &   $\simeq1.2\,\mathrm{R}_\odot $  \\
Ellipticity                            &   0.021                  \\
\hline                                                                  
\end{tabular}

\clearpage

\centerline{\textbf{\large Supporting Online Material}}

\section{A dynamical model for the emission of a rotating gas ring}

The SDSS spectroscopy of the white dwarf SDSS1228+1040 presented in
the main paper gave reasons to believe that the Ca\,II emission line
profiles observed in the $I-$band are are from a circumstellar gas
disk. Here we outline the model we developed to fit the Ca\,II line
profiles.  It is apparent that there must be some form of asymmetry at
work as the peaks of the line profiles are asymmetrical. The shape is
very reminiscent of the ``V/R'' (violet/red) asymmetries seen in the
emission line profiles of B-type emission line stars \textit{(S1--S3)}
\nocite{okazaki91-1, meillandetal06-1, stee+dearaujo94-1}. These are
ascribed to a one-armed spiral wave which can be approximately
modelled with a series of elliptical orbits.  Drawing upon this
analogy, we modelled the profiles of SDSS\,1228+1040 by assuming that
the orbits in the disk took the form of a series of co-aligned
elliptical orbits of identical eccentricity. We further allowed the
emission-line flux per unit area from the disk to vary as $1 +
\epsilon \cos \theta$ where $\theta$ is the angle from periastron and
$\epsilon$ is a measure of the asymmetry. A less obvious feature of
the emission line profiles of Fig.\,3 is that they are fainter at line
centre relative to the double-peaks than can be explained by
optically-thin line emission. That they are optically thick is also
strongly suggested by the relatively equal line strengths within the
Ca\,II triplet, even though Ca\,II\,850 nm has only one tenth the
oscillator strength of Ca\,II\,854 nm. We therefore turned to the
formalism of Horne \& Marsh \textit{(S4)}\nocite{horne+marsh86-1} for optically thick
line formation in which emission from any point in the disk takes on
an azimuthal asymmetry owing to the local velocity shear, with
strongest emission along the four directions at $45^\circ$ to the
radial direction and weakest emission in the radial direction and in
the direction of gas flow.  This has the effect of deepening the
central dip between the line peaks. We had to modify the prescription
of Horne \& Marsh \textit{(S4)}\nocite{horne+marsh86-1} slightly to allow for
elliptical as opposed to circular motion, although the effect overall
was weak, since the ellipticity implied by the data was small.

Apart from an arbitrary normalisation factor, the following
parameters were needed to specify the model fully: (a) $M_\mathrm{W} = 0.75$,
the mass of white dwarf which determines the relation between orbital
radius and speed, (b) $e = 0.021$, the eccentricity, (c) the angle of
the periastron to our line-of-sight, defined so that at $0^\circ$ we
look along the major-axes from peri- towards apastron, while at
$90^\circ$ we look along the minor axis with the periastron emission
red-shifted, (d) inner semi-major axis $a_\mathrm{in} =
0.64\,\mathrm{R}_\odot$, (e) outer semi-major axis $a_\mathrm{out} =
1.2\,\mathrm{R}_\odot$, (f) exponent of the radial power law used to
set the emissivity $\propto r^\alpha$ where $\alpha = - 0.9$, (g) an
optical depth parameter $Q' = 2.5$, analogous to the combination $Q
\sin i \tan i$ from Horne \& Marsh \textit{(S4)}\nocite{horne+marsh86-1}. Since $Q \sim 1$, this
suggests that the orbital inclination $i \sim 70^\circ$, and finally
(h) the azimuthal asymmetry factor mentioned above with value $\epsilon = -0.19$.
All parameters other than the mass of the white dwarf were allowed to
vary. Experience from similar line profiles in cataclysmic variable
stars suggests that systematic effects rather than statistical errors
dominate, and the fit in this case, while qualitatively reasonable, is
not a statistically good match to the data. Therefore we refrain from
giving formal confidence intervals, but instead discuss which features
in the data lead to constraints on these parameters to help the reader 
evaluate our model.

Apart from the allowance for optical depth, we made no allowance for
inclination because its only effect is to scale the velocities in
proportion to $\sin i$. This scaling can be counter-balanced by
re-scaling the major-axis parameters $a_\mathrm{in}$ and $a_\mathrm{out}$ such that
$a \propto \sin^2 i$. Thus the value of the outer semi-major axis
$a_{out} = 1.2\,\mathrm{R}_\odot$,
which is almost the outer radius since the eccentricity is small, is an upper limit to the real value.  Since
the optical depth parameter (which comes from the depth of central dip
in the profiles) points to a large inclination, we believe this to be
close to the true value, with a likely range of order $0.9$ to
$1.2\,\mathrm{R}_\odot$; this is quite robust as the outer major-axis
limit is set by the velocities of the line peaks which are
well-defined. An interesting feature of our model is that we also
require a fairly large inner semi-major axis $a_{in} =
0.64\,\mathrm{R}_\odot$, which is subject to the same inclination
uncertainties as $a_{in}$, but also additional uncertainty owing to
correlation with parameter (f), the emissivity power law exponent. The
inner cut-off is required to fit the steep drop-off in the line wings;
this presumably reflects an absence of emission rather than an absence
of material given the presence of metals in the white dwarf's
photosphere. It could be caused, for instance, by ionisation of the
Ca\,II ions close to the white dwarf.

The eccentricity, periastron angle and azimuthal asymmetry parameters
are necessary because of the asymmetry of the profiles. The exact
nature of the asymmetry is uncertain and thus we regard these
parameters as the least reliable, although there is no doubt of the
asymmetry. If our interpretation in terms of eccentric orbits
contributing to a one-armed spiral is correct, further observations of
the object are of considerable interest for we expect such orbits to
precess.  The asymmetries in Be stars precess within the distorted
gravitational fields of the rapidly rotating stars for instance on
decade-long timescales. The narrow photospheric Mg\,II absorption line
shows that SDSS\,1228+1040 is not rapidly rotating, but general
relativity alone implies a precession rate of order $5^\circ$ per year
in the orientation of the outermost elliptical orbits, and pressure
effects within the disk could contribute as well and may indeed be
much more significant \textit{(S5)}\nocite{ogilvie01-1}.  Such precession effects
must be computed within the framework of fluid disk models since the
precession rate is a strong function of radius and would be expected
to randomise the orbit orientations if fluid effects were not taken
into account; the Be stars are proof that it is possible for similar
asymmetries to persist over long timescales (years to decades).

Our data do not directly constrain the thickness of the disk, but we
can argue that it must be thin as follows: LTE (local thermodynamic
equilibrium) models of the disk suggest a temperature of around $4500$
to $5500\,\mathrm{K}$ for the disk. Any hotter and numerous metal
lines that are not observed start to appear, while any colder and the
Fe\,II lines disappear. The sound speed, assuming dominance by CNO
elements in the absence of hydrogen and helium, is of order $C_S =
2\,\mathrm{km}\,\mathrm{s}^{-1}$. The thickness of the disk is then of
order $H = (C_S/V_\mathrm{Orb}) R$ where $V_\mathrm{Orb}$ is the
orbital velocity at radius $R$ in the disk \textit{(S6)}\cite{pringle81-1}. This
works out at $\sim 0.005 \,\mathrm{R}_\odot$, comparable to the size
of the white dwarf. The same LTE models place a weak lower limit on
the Ca\,II/H ratio which must be $> 3$ times solar; the data in hand
are consistent with no hydrogen at all.

\section{An investigation of white dwarfs in the SDSS}
We have investigated all white dwarfs with hydrogen-dominated
atmospheres brighter than $g=17.5$ contained in Data Release 4 of the
Sloan Digital Sky Survey (SDSS) \textit{(S7)}\cite{eisensteinetal06-1} for signs of
either photospheric metal absorption lines or Ca\,II emission in the
$I$-band. Because the SDSS spectroscopy has a relatively low spectral
resolution ($\lambda/\Delta\lambda\simeq1800$), the detection of metal
lines is limited to relatively high abundances such as observed in
SDSS\,1228+1040. We have not identified any additional white dwarf
which exhibits significant metal absorption. The detection of Ca\,II
emission with equivalent width comparable to the lines detected in
SDSS\,1228+1040 is robust for spectra in the considered magnitude
range. However, we have identified only one additional white dwarf
which shows flux excess in the region of the Ca\,II triplet
(SDSS\,J104341.53+085558.2, Fig.\,S1). The equivalent width of the
Ca\,II triplet in SDSS\,1043+0855 is a factor three lower than in
SDSS\,1228+1040, and its SDSS spectrum suffers from strong residuals
from imperfect night sky line subtraction in the $I$-band. A spectral
fit to the SDSS spectrum of SDSS\,1043+0855 results in a temperature
of $18900\pm300$\,K and a surface gravity of $\log g=8.2\pm0.1$,
similar to the parameters found for SDSS\,1228+1040.

Circumstellar metal disks are most likely to be found around white
dwarfs with substantial metal abundances. Published spectroscopy of
known metal-polluted white dwarfs with temperatures in excess of
15\,000\,K does typically not cover the red end of the optical
wavelength range (e.g. \textit{S8}\nocite{zuckermanetal03-1}), and an assessment of
the frequency of circumstellar gas disks will require a systematic
survey of these stars. 

\bibliographystyle{Science}

\clearpage

\centerline{\includegraphics[angle=-90,width=15cm]{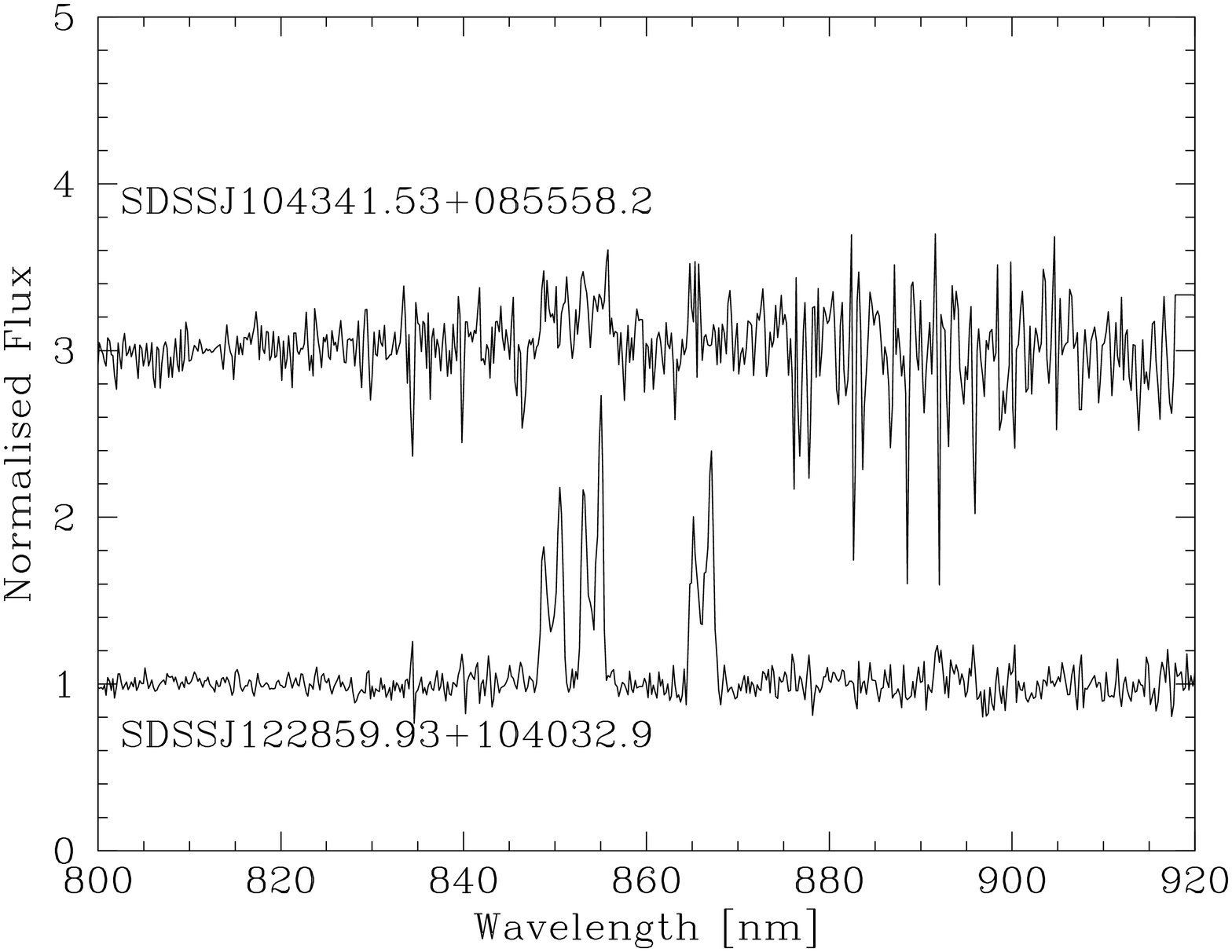}}

\bigskip
\noindent
\textbf{Supporting Figure\,1.} Normalised spectra of SDSS\,1228+1040
and SDSS\,1043+0855, the only other white dwarf among the brightest
406 hydrogen-dominated white dwarfs in Data Release 4 of the SDSS that
exhibits excess emission in the region of the Ca\,II triplet. The
spectrum of SDSS\,1043+0855 has been offset by two units. The large
amount of noise in the spectrum of SDSS\,1043+0855 is due to
imperfections in the removal of strong night sky lines. 

\end{document}